\documentstyle[epsf]{aipproc}

\begin{document}
\title{Did GRB 970228 and GRB 970508 present similar optical properties?}

\author{C. Bartolini$^1$, G.M. Beskin$^2$, A. Guarnieri$^1$,\\
N. Masetti$^1$ \& A. Piccioni$^1$}
\address{$^1$Dipartimento di Astronomia, Universit\`a di Bologna, Italy\\
$^2$ Special Astrophysical Observatory of RAS, Nizhnij Arkhyz, Russia}

\maketitle

\begin{abstract}
The analysis of the light curves of the optical counterparts of GRB 970228
and GRB 970508 points out remarkable similarities.
The spectral distribution obtained from the color indices shows 
that both the transients became bluer during the increasing stage, 
and redder after the maximum. 
A main difference concerns the behaviour of the optical fading which is well
fitted by a single power law in the case of GRB 970508 but not in the case of 
GRB 970228.
\end{abstract}

\section*{GRB 970228}

The optical counterpart of GRB 970228 was discovered by van Paradijs et al.
[18] who observed it for the first time on February 28.99 UT, 1997. 
Early observations by Guarnieri et al. [11] in the $B$ and $R$ bands 
on February 28.83 allowed the detection of
the presence of a rising branch in the light curve. 

\begin{figure}[t!] 
\vspace{80mm}
\caption{{\bf Left}: $R$ light curve of the optical transient associated 
with GRB 970228. Data were taken from literature (see text),
$t$ is the time interval, in days, 
since the $\gamma$--burst. 
{\bf Right}: broadband spectra of 
GRB 970228; the flux densities were
computed from the photometric data by Guarnieri et al. [11] (triangles), 
van Paradijs et al. [18] (squares) and Sahu et al. [17] (circles).
The HST data were multiplied by a factor of 10 for sake of clarity.
The data corresponding to the open symbols are evaluated from the measured 
values (filled symbols) using a linear interpolation of the flux densities.}
\end{figure}

Fig.1 (left) shows the $R$ light curve obtained by using 
the collection of observations reported by Galama et al. [8] and the
data by Guarnieri et al. [11] and Fruchter et al. [6].
Pedichini et al.'s [15] observations were not included because of the 
difficulty to reduce their color system to the $R$ band.
A fitting of the optical data with a single power
law yields an index $\alpha_{\rm opt}=1.21\pm0.02$; however, 
data reported by 
Galama et al. [8] show a decay behaviour which seems to follow a power
law with spectral index $\alpha_{\rm opt}=2.1$ before March 6, 1997, and
with $\alpha_{\rm opt}<0.35$ after that day.

The evolution of the optical spectrum of GRB 970228 is presented in Fig. 1
(right).
The flux densities were evaluated from the photometric data taken on 
February 28.83 [11], February 28.99 [18], March 26.4 and April 7.2
[17] using the formulae by 
Fukugita et al. [7]. The optical transient became bluer 
during the rise and then, on its way to quiescence, 
significantly reddened, going from $B-V=0.5$ and $V-R=0.5$ on February 28.83 
to $V-R=0.4$ and $V-I=0.7$ on February 28.99, to $V-R=0.9$ and $V-I=1.9$ on 
March 26.4. 

\section*{GRB 970508}

Surprisingly --- but not too much in the light of the behaviour found by 
Guarnieri et al. [11] for GRB 970228 --- the optical transient discovered
by Bond [1] inside the GRB 970508 error box
increased in brightness during the first two days after the $\gamma$--burst,
reaching the peak around May 10.8 UT. According to Castro--Tirado et al. 
[2] during this phase the object was very blue. 

Its $R$ light curve is shown in Fig. 2 (left). 
The data were taken from the IAU Circulars, from several papers published in
these proceedings and from Kelemen [12]. An upper limit to the 
brightness was obtained with the Bologna 
Telescope on May 23.89 UT.
A power--law decay fits the data until 50--60 days after the
$\gamma$--burst. We obtain 
$\alpha_{\rm opt}=1.34\pm0.02$, which is slightly higher than the values
found by Djorgovski et al. [3], Kopylov [13] and Fruchter et al. [5].
Then, the light curve seems to flatten about at $R\approx25.2$, which
might be the magnitude of the host galaxy [16].

\begin{figure}[t!] 
\vspace{80mm}
\caption{
{\bf Left}: $R$ light curve of the optical transient associated with
GRB 970508. Data and upper limits are from various authors (see text for 
details). $t$ represents the time interval, in days, since the 
$\gamma$--burst occurred. {\bf Right}: broadband spectra of 
GRB 970508 built with the flux densities
computed from the photometric data by Galama et al. [9] on May 10.05
(triangles) and on 11.03 (squares), and by Groot et al. [10] on 12.03 
(circles).
}
\end{figure}

The observations by Galama et al. [9], Groot et al. [10], Kopylov et al. [14] 
and Kopylov [13] are 
homogeneous and cover almost all the optical band in the same time 
lapse. From these data, we evaluated the flux densities and the broadband 
spectral evolution of GRB 970508 using the same procedure applied to GRB 970228 
data; the results are reported in Figs. 2 (right) and 3
(left). On May 10.05 the object
was red; then, it became progressively bluer on May 10.77, 10.93, until it 
reached a maximum on May 11.03, close to the light peak.
On May 11.76, the spectrum became flatter (Fig. 3, left) but 
on May 12.03, the slope rose again at short wavelengths
(Fig. 2, right). Subsequently 
the spectrum started to redden,
peaking in the $R$ on May 22.00. This behaviour is confirmed 
by the trend of the color of the optical transient (Fig. 3, right).

\begin{figure}[t!] 
\vspace{80mm}
\caption{
{\bf Left}: broadband spectra of GRB 970508 built with the flux densities
computed from the photometric data by Kopylov et al. [14] and Kopylov [13].
The upper limit in the figure refers to May 22.00.
{\bf Right}: $B-V$ colors of GRB 970508 from May 10 to May 14.
Dutch data are indicated with open squares, and Russian ones with filled
squares. 
}
\end{figure}

It is noteworthy that the $B-V$ secondary maximum occurred around the 
appearance of a transient radio emission associated to GRB 970508. 
Indeed, nearly a week after the onset of the $\gamma$--burst,
Frail et al. [4] found a flaring radio source, increasing in 
brightness, which was coincident with the positions of the X--ray and optical 
counterparts of GRB 970508.

\section*{Comparison between the light curves}

Although the light curve of GRB 970228 is less sampled, 
we can see remarkable similarities with that of GRB 970508, 
by comparing the left panels of Figs. 1 and 2.
Both light curves show a rising branch.
In particular, the maximum occurred 
$\sim$1 day after the burst for GRB 970228, and around 2 days
after for GRB 970508. 
Therefore, the presence of a delayed maximum in the optical light curve appears 
to be a common feature for all the GRBs observed thus far.

The two light curves show similar overall decays. But if the $R$ light curve 
of GRB 970508 can
be fitted with a single power law until 60 days from the onset of the 
$\gamma$--burst, the one of GRB 970228 does not;
indeed, Guarnieri et al. [11]
noticed a more rapid decrease during the first 3 days after the $\gamma$ event.

\section*{Comparison between the spectra}

The spectral distribution (Figs. 1 and 2, right, and Fig. 3, left)
and the $B-V$ color index (Fig. 3, right), 
show that both the optical transients became bluer during the rising phase,
and then reddened during their approach to quiescence.
This could be simply understood in the light of the `fireball' model
(e.g. Wijers et al. [19]): a moving shock wave heats the surrounding
medium, which then cools down. In this framework, the fast spectral and color 
variation of GRB 970508 in the optical is therefore surprising.

Due to the paucity of the observations, we cannot say if GRB 970228 showed a 
similar behaviour during the first days after the burst. However, if these
variations are connected with the radio transient and since no radio emission
was revealed for GRB 970228, it is possible that no such changes in the
optical spectral energy distribution took place in GRB 970228.

Another explanation for these spectral changes could arise from the presence of
cool absorbing material placed on the line--of--sight at a distance of about 2 
or 3 light--days from the center of the burst; it could have produced a `dip' 
at shorter wavelengths suddenly after the light peak, and then could have been 
shocked by the blast wave in the following days.

It seems that GRB 970228 was redder than GRB 970508 at 
light peak ($B-V=+0.5$ instead of +0.1). However we are not sure that the 
minimum $B-V$ for GRB 970228 has been observed. Moreover, this value could 
be affected by interstellar (and/or intergalactic) 
absorption; indeed, GRB 970228 lies closer to the galactic plane than GRB 
970508. Actually, 
the absorption in the $V$ towards GRB 970228 is $A_V = 0.4$ mag [18],
while towards GRB 970508 is only $A_V = 0.08$ mag 
(derived from the $A_B$ value reported by Djorgovski et al. [3]).

\section*{Acknowledgements}
This investigation is supported by the University of Bologna (Funds for 
selected research topics). We thank G. Valentini for several 
useful discussions.

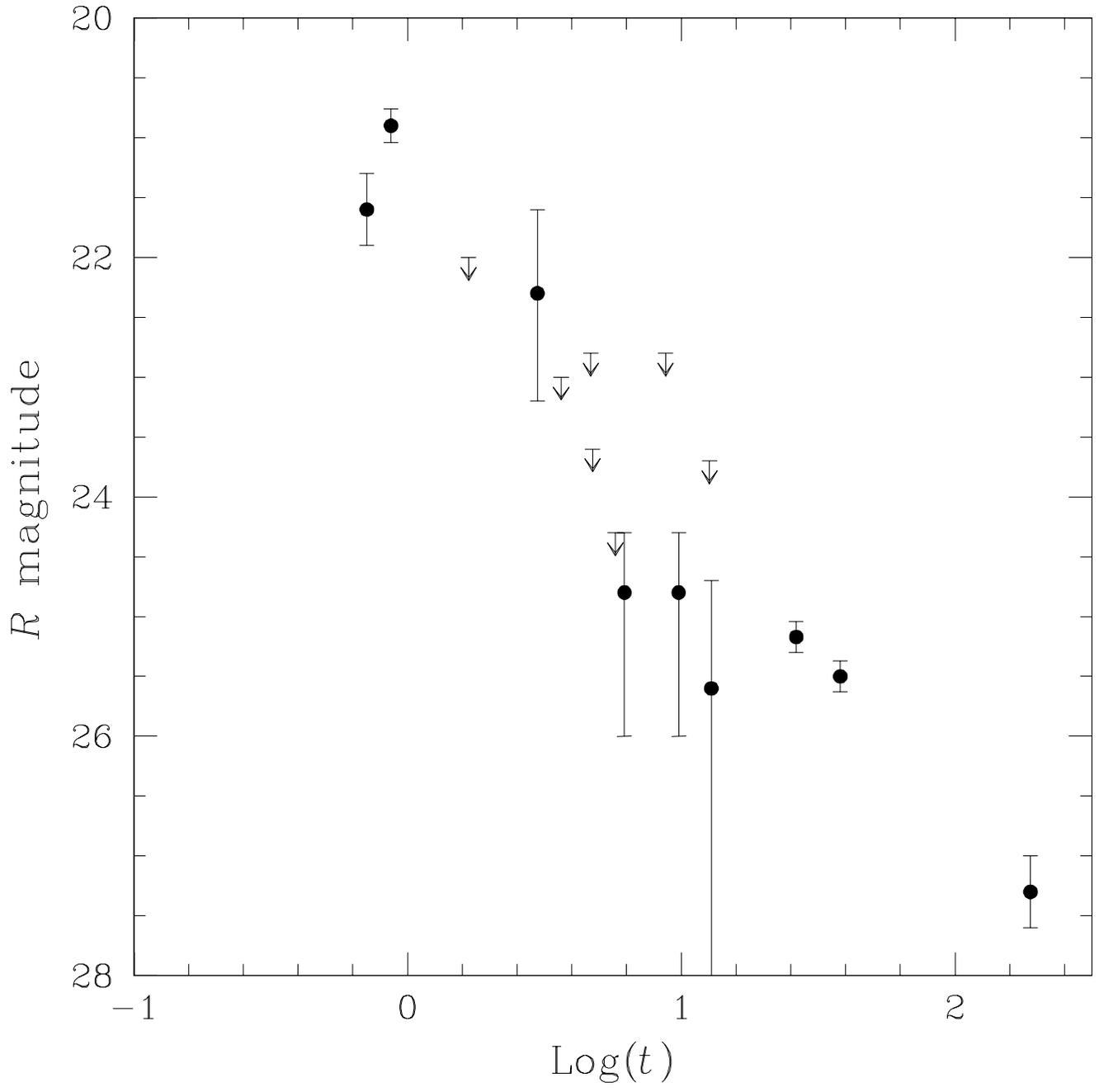
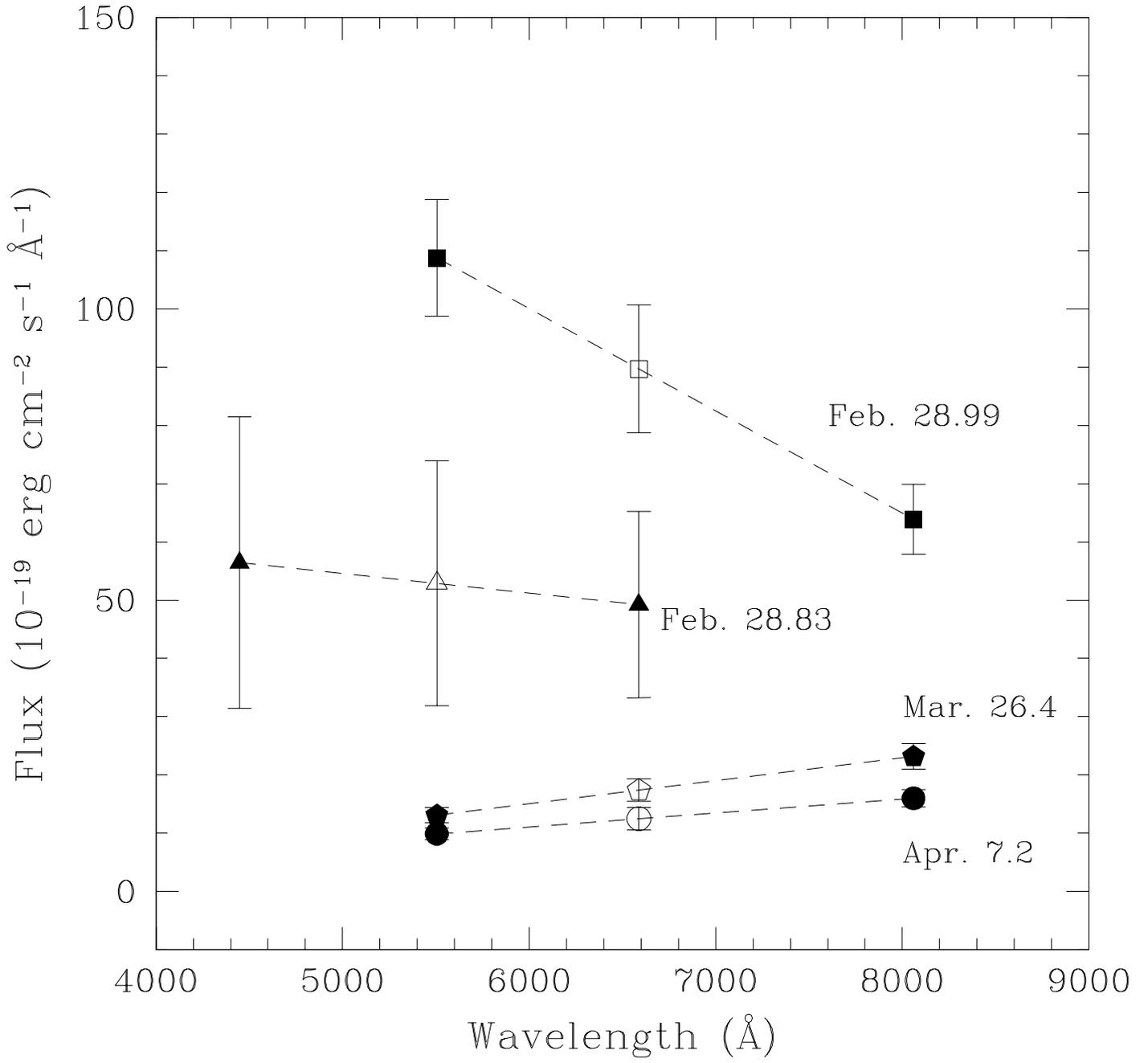
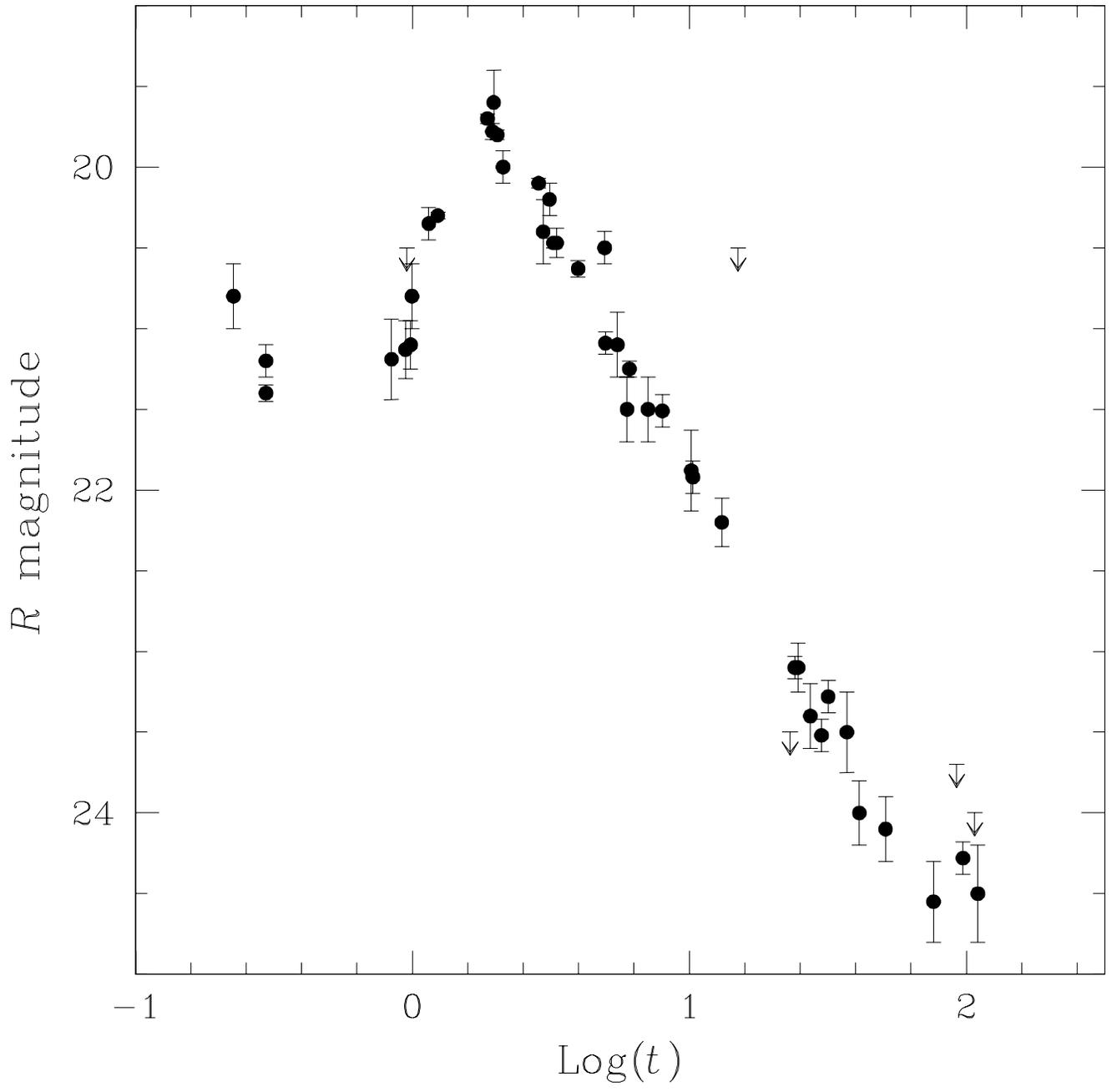
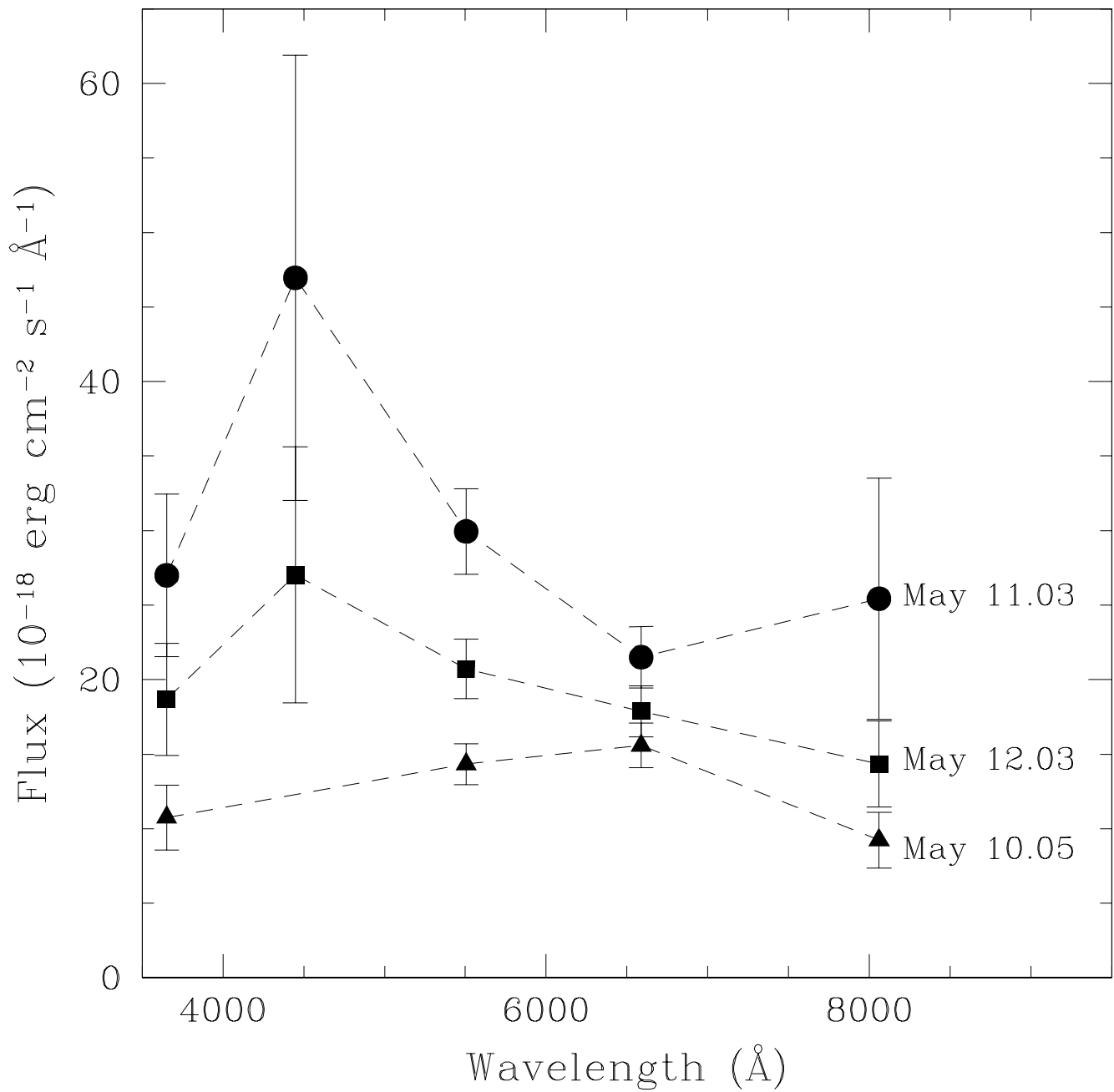
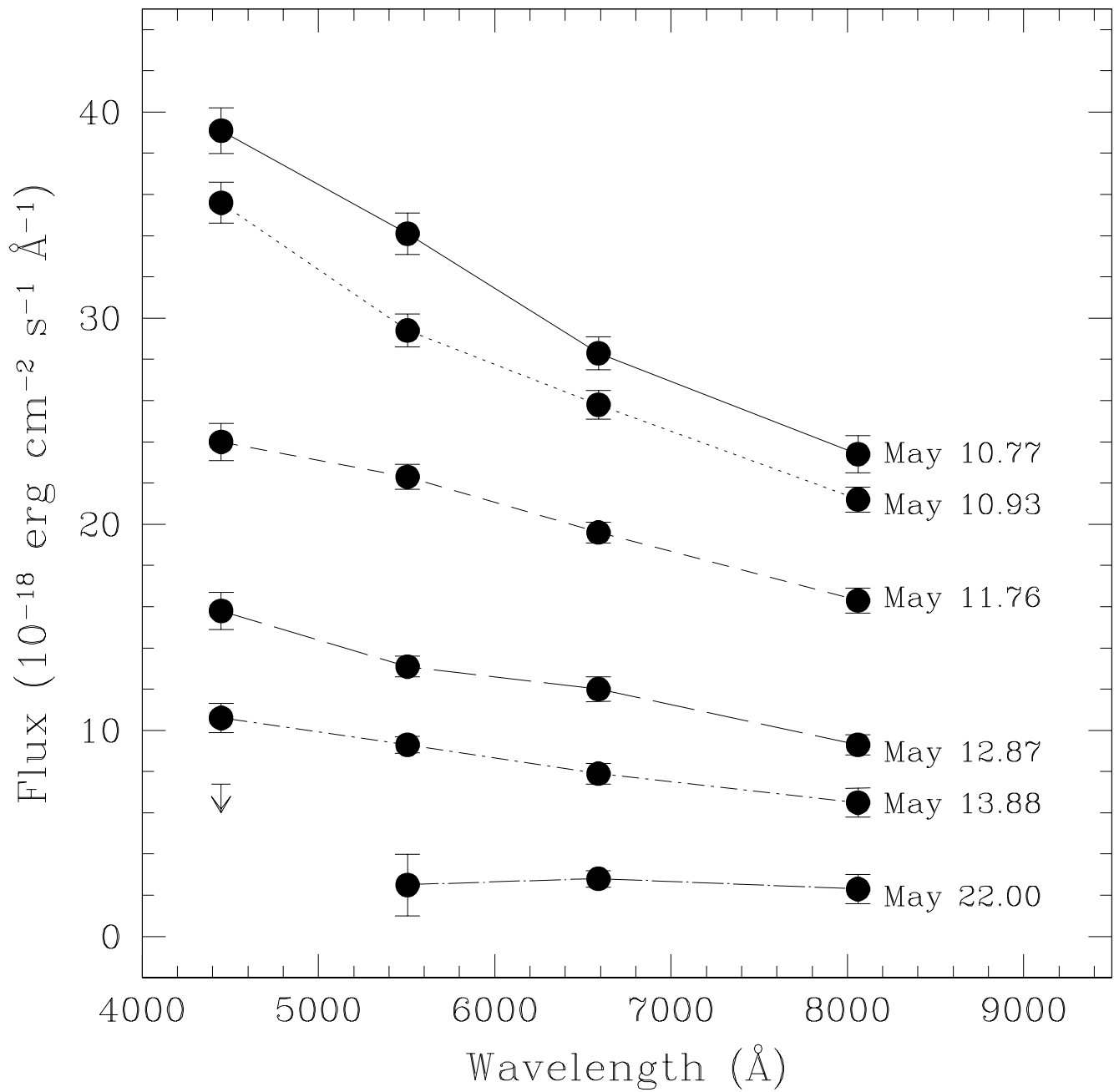
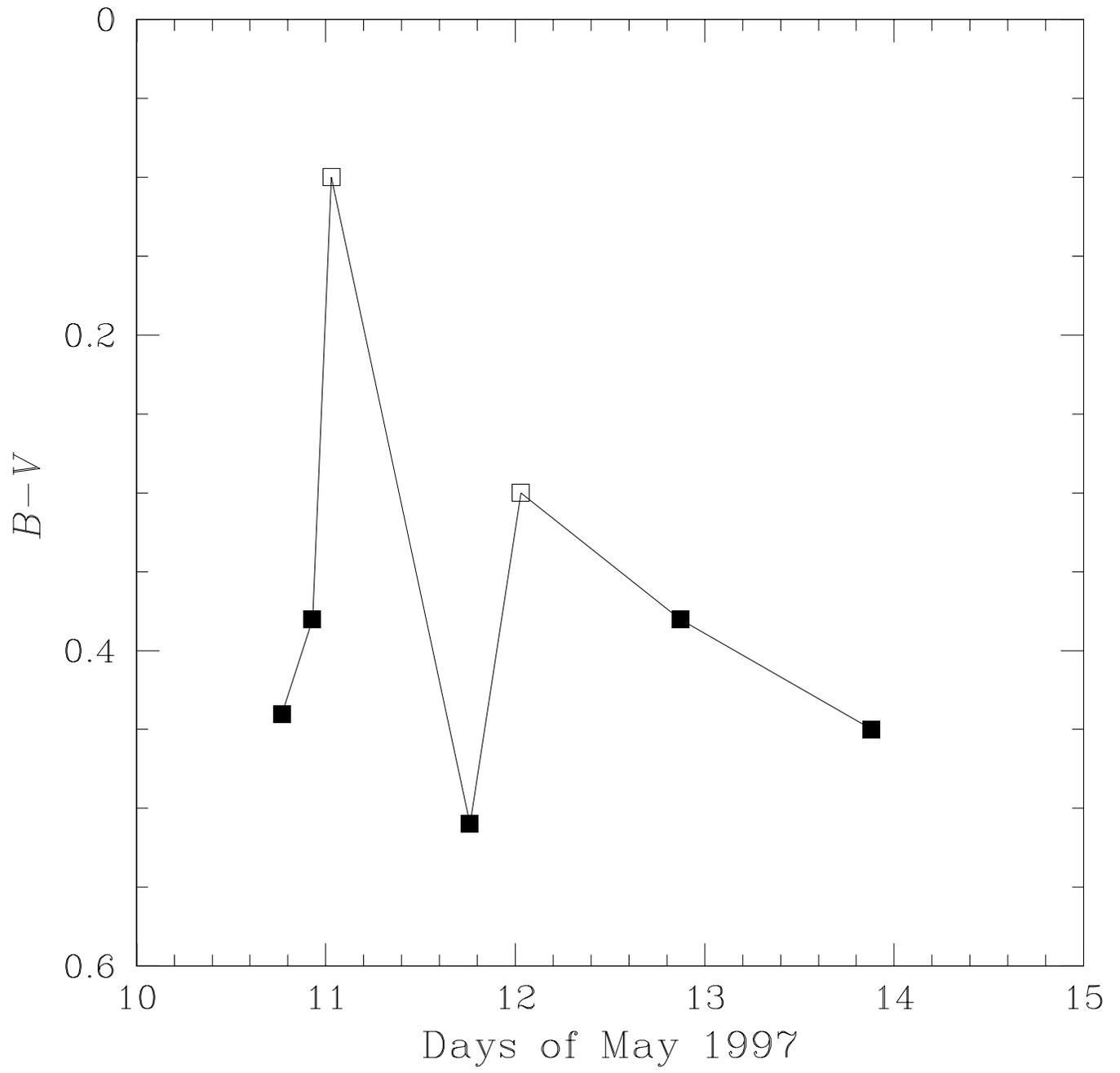
\end{document}